
\magnification 1200

\input epsf.tex

\newcount\figno
\figno=0
\def\fig#1#2#3{
\par\begingroup\parindent=0pt\leftskip=1cm\rightskip=1cm\parindent=0pt
\baselineskip=11pt \global\advance\figno by 1 \midinsert
\epsfxsize=#3 \centerline{\epsfbox{#2}} \vskip 12pt
#1\par
\endinsert\endgroup\par}
\def\figlabel#1{\xdef#1{\the\figno}}

\baselineskip=16pt

\font\twelvebf=cmbx12

\def\n{\noindent}
\def\[{[\![}
\def\]{]\!]}

\def\s{\smallskip}



\def\t{\theta}

\def\a{\alpha}
\def\b{\beta}
\def\g{\gamma}
\def\d{\delta}

\def\ra{\rangle}

\bigskip\n

\vfill\eject


\vskip 7mm

\centerline {\twelvebf A Note on the Statistics of Hardcore Fermions}



\leftskip 50pt
\vskip 32pt
\noindent
T D Palev\footnote{$^{a)}$}{E-mail: tpalev@inrne.bas.bg}

\noindent
Institute for Nuclear Research and Nuclear Energy, 1784 Sofia,
Bulgaria

\bigskip\bigskip

\noindent {\bf Abstract.} It is shown that the statistics of the hardcore
fermions is $A-$ superstatistics of order one [see T.D.P. {\it J. Math. Phys.}
{\bf 21} 1293 (1980)]. The Pauli principle for these particles is formulated.
The Hubbard operators, which constitute a basis in the Lie superalgebra
$gl(1|n)$, are expressed via the creation and the annihilation operators of the
hardcore fermions.

\bigskip
\n




\vfill\eject \leftskip 0pt \vskip 48pt

\vskip 12pt

\bigskip\noindent

\bigskip
\bigskip

The aim of the present note is to show that the so-called hardcore
fermions can be viewed as particles obeying the $A$-superstatistics
introduced in [1] (see also the review paper [2]).

Hardcore fermions appear implicitly in various Hubbard lattice models [3] of
strongly correlated electron systems and in particular in models of high
temperature superconductivity. By definition the particles are hardcore if they
obey the hardcore restriction (HC restriction): each site of the lattice cannot
accommodate more than one particle.

The HC restriction is stronger than the Pauli principle. Indeed, if the number of the
orbitals at site $i$ is $n_i$, then the Pauli principle for fermions asserts
that the site $i$ can accommodate up to $n_i$ particles, whereas the HC
property admits at most one particle.

Physically, the HC property stems from the repulsion between the
electrons at low energies. Mathematically, this property is
described by projecting down the entire fermionic Fock space $W$
onto the subspace $W^{(1)}$ of states with at most one particle per
site. The latter can be achieved in different ways. Some of the
pioneering papers to mention are the Gutzwiller variational method
[4], further developed by Kotliar and Ruckenstein [5], the
slave-boson approach of Barnes [6], extended by Coleman [7], Read
and Newns [8], but in general the literature on the subject is vast.

We proceed to show that the statistics of hardcore particles is
$A-$superstatistics of order 1 [1]. Consider a lattice with $N$ sites.
Let
$$
F(f_{11}^\pm ,f_{12}^\pm,...,f_{1n}^\pm;~ f_{21}^\pm
,f_{22}^\pm,...,f_{2n}^\pm;...;f_{N1}^\pm ,f_{N2}^\pm,...,f_{Nn}^\pm) \eqno(1)
$$
be any polynomial of $Nn$ pairs of Fermi creation and annihilation operators
(CAO's), where $f_{i \a}^\pm$ creates/annihilates a fermion at the site
$i=1,2,...,N,$ with, say, a flavor index (including spin and other internal
characteristics) $\a=1,2,...,n$. Denote by ${\cal P}$ a projection operator
from the entire fermionic state space $W$ onto the subspace $W^{(1)}$ defined
above. Then
$$
\eqalign{ & {\cal P}F(f_{11}^\pm ,f_{12}^\pm,...,f_{1n}^\pm;~ f_{21}^\pm
,f_{22}^\pm,...,f_{2n}^\pm;...;f_{N1}^\pm ,f_{N2}^\pm,...,f_{Nn}^\pm){\cal
P}\cr & =F(a_{11}^\pm ,a_{12}^\pm,...,a_{1n}^\pm;~ a_{21}^\pm,
a_{22}^\pm,...,a_{2n}^\pm;...;a_{N1}^\pm ,a_{N2}^\pm,...,a_{Nn}^\pm),\cr }
\eqno(2)
$$
where
$$
a_{i\a}^+ ={\cal P}f_{i\a}^+{\cal P},~~~a_{i\a}^- ={\cal
P}f_{i\a}^-{\cal P},~~ i=1,2,..,N,~~\a=1,2,...,n  \eqno(3)
$$
For instance, if
$$
H=-t\sum_{ij}\sum_\a (f_{i\a}^+ f_{j\a}^- + f_{j\a}^+ f_{i\a}^-)+
U\sum_i~\sum_{\a\ne \b} f_{i\a}^+ f_{i\a}^-  f_{i\b}^+ f_{i\b}^-, \eqno(4)
$$
then
$$
{\cal H}={\cal P}H{\cal P}= -t\sum_{ij}\sum_\a (a_{i\a}^+ a_{j\a}^- + a_{j\a}^+
a_{i\a}^-)+ U\sum_i~\sum_{\a\ne \b}a_{i\a}^+ a_{i\a}^- a_{i\b}^+ a_{i\b}^-.
\eqno(5)
$$
(some of the ${\cal P}$'s in (3) can be skipped, but  we keep them for symmetry).

By a straightforward computation, one verifies that at each site
$i=1,2,..,N,$ the above operators satisfy the following relations in
$W^{(1)}$:
$$
\eqalignno { & [\{a_{i\a}^+,a_{i\b}^-\},a_{i\g}^+] =~~\delta_{\b\g}a_{i\a}^+ -
\delta_{\a\b}a_{i\g}^+, & \cr &
[\{a_{i\a}^+,a_{i\b}^-\},a_{i\g}^-]=-\delta_{\a\g}a_{i\b}^- +
    \delta_{\a\b}a_{i\g}^-, & (6)\cr
&   \{a_{i\a}^+,a_{i\b}^+\}=
    \{a_{i\a}^-,a_{i\b}^-\}=0. &  \cr
}
$$
The triple relations (6) are defining relations for the creation and
annihilation operators of $A-$superstatistics at site $i$ [1].

At different sites, the operators anticommute,
$$
\{a_{i\a}^+,a_{j\b}^+\}=\{a_{i\a}^+,a_{j\b}^-\}=\{a_{i\a}^-,
a_{j\b}^-\}=0,~~i\ne j=1,...,N.   \eqno(7)
$$
From (6) one concludes that if the creation and the annihilation operators
$a_{i\a}^\pm$ are postulated to be odd elements, then the linear span
$$
lin.span.\{ a_{i\a}^\pm,~\{a_{i\b}^+,a_{i\g}^-\}|\a, \b, \g
=1,...,n \} \eqno(8)
$$
is a Lie superalgebra (LS) with an even subalgebra specified by
$$
lin. span.\{ \{a_{i\b}^+,a_{i\g}^-\}|\a, \b, \g =1,...,n \}.
$$
A more detailed analysis [1] shows that at each site $i$, the
operators $a_{i\a}^\pm$ generate the Lie superalgebra
$sl(1|n)^{(i)}$. Then (7) implies that the operators $a_{i\a}^\pm$,
$\a=1,2,...,,n$, $i=1,2,...,N,$ generate a Lie superalgebra which is
a direct sum of $N$ identical copies of $sl(1|n)$,
$$
A(N,n)= sl(1|n)^{(1)}\oplus sl(1|n)^{(2)}\oplus ...\oplus
sl(1|n)^{(N)}. \eqno(9)
$$

The circumstance that the hardcore CAO's generate a Lie superalgebra carries
important information. The immediate conclusion is that the hardcore operators
give one particular solution of the relations (6) or - with another words -
these CAO's give one particular representation of each Lie superalgebra
$sl(1|n)^{(i)}$ and together with (7) - a representation of the LS (9). Next,
it is known that the LS $sl(1|n)^{(i)}$ has several other representations,
i.e., several other solutions of (6) and in view of (7) also the algebra
$A(N,n)$ has different solutions. Are these new representations of any
interest? What is their physical interpretation, if any? These are the
questions we will address next.

As usually we shall write $sl(1|n)^{(i)}$  in a basis of the general linear Lie
superalgebra $gl(1|n)^{(i)}$. Such an extension is convenient, since the
$gl(1|n)^{(i)}$ basis is simpler. Moreover it does not change anything related
to $sl(1|n)^{(i)}$ since the representation space remains the same (every
irreducible $sl(1|n)^{(i)}$ module can be extended to an irreducible
$gl(1|n)^{(i)}$ module). But as we shall see almost immediately, the extension
is more than simply convenient.

As a basis in $gl(1|n)^{(i)}$ we choose $(n+1)^2$ generators $X_{AB}^{(i)}$
with $A,B=0,1,...,n$. The odd generators are $X_{0\a}^{(i)}$ and $X_{\a
0}^{(i)}$, $\a=1,2,..,n$. All other generators are even. The $X$ operators satisfy the
supercommutation relations:
$$
[X_{AB}^{(i)},X_{CD}^{(i)}]_\pm=\delta_{BC}X_{AD}^{(i)}\pm
\delta_{AD}X_{CB}^{(i)}, \eqno(10)
$$
whereas at different sites
$$
[X_{AB}^{(i)},X_{CD}^{(j)}]_\pm=0, ~~i\ne j. \eqno(11)
$$
In the above $A,B,C,D=0,1,...,n$, and the upper sign $(+)$ stands for the case
when both generators in the LHS are odd, otherwise the lower sign $(-)$ should
be adopted.

The supercommutation relations (10) determine completely the LS
$gl(1|n)^{(i)}$. Here they are written in somewhat unusual for this LS form. We
have adopted such notation because the $X$ operators (10), called Hubbard
operators, play an important role in condensed mather physics as an alternative
way for description of strongly correlated electron systems.

The Hubbard operators yield one possible basis of the LS $gl(1|n)^{(i)}$. In
fact these operators define a particular representation, the fundamental
$(n+1)-$dimensional representation of each $gl(1|n)^{(i)}$. In a matrix form
the Hubbard operators are nothing but the $(n+1)-$dimensional matrix units
($X_{A,B}^{(i)}$ has 1 at position (A, B) and 0 elsewhere).

Locally, at each  site $i$, the creation and annihilation operators
$a_{i\a}^\pm$ together with $X_{00}^{(i)}$ generate $gl(1|n)^{(i)}$. Therefore
the Hubbard operators can be expressed via the hardcore creation and
annihilation operators and $X_{00}^{(i)}$:
$$
X_{0\a}^{(i)}=a_{i\a}^-,~~X_{\a 0}^{(i)}=a_{i\a}^+, ~~\a=1,...,n. \eqno(12)
$$
Observe that the CAO's coincide with the odd $X-$operators. Then
$$
X_{\a\b}^{(i)}=\{a_{i\a}^+,a_{i\b}^-\}, ~~\a\ne \b=1,...,n,
\eqno(13a)
$$
$$
X_{\a\a}^{(i)}=\{a_{i\a}^+,a_{i\a}^-\}-X_{00}^{(i)},~~\a=1,...,n.\eqno(13b)
$$
Below, see (21), we express also $X_{00}^{(i)}$ via the CAO's, so that all
Hubbard generators become functions of only CAO's.

In [1] we have introduced a concept of Fock representations of a
simple Lie (super)algebra. The $sl(1|n)$  Fock modules, considered
here, are finite-dimensional and irreducible. They are labelled by
all positive integers $p=1,2,...$, the order of statistics. As in
parastatstics the representation space $W(n,p,i)$ at site $i$ and
with order of statistics $p$ is reconstructed from the relations
$$
a_{i\a}^- a_{i\b}^+ |0\ra=\d_{\a\b}p|0\ra, ~~~a_{i\a}^- |0\ra= 0.
\eqno(14).
$$
Without loss of generality for $sl(1|n)$, we extend $W(n,p,i)$ to a $gl(1|n)$
module setting
$$
X_{00}^{(i)}|0\ra =p |0\ra. \eqno(15)
$$
The representations corresponding to different orders of statistics $p$ are
inequivalent finite-dimensional irreducible representations.
At each site $i$ all states
$$
|p;\t_{i1},\t_{i2},...,\t_{in}\ra={\sqrt{(p-\sum_\a
\t_{i\a})!\over{p!}}}
 (a_{i1}^+)^{\t_{i1}}...(a_{in}^+)^{\t_{in}}|0\ra,~~
\sum_{\a=1}^n \t_{i\a} \leq \min(n,p) \eqno(16)
$$
with $\t_{i1},...,\t_{in}=0,1$, constitute an orthogonal  basis
in $W(n,p,i)$.

The transformations of the basis under the action of the odd generators read
[1]:
$$
\eqalignno{
& a_{i\a}^-|...,\t_{i\a},..\ra = \t_{i\a}(-1)^{\t_{i1}+..+\t_{i,\a -1}}
 \sqrt{p-\sum_{\b} \t_{i\b} +1} |...,\t_{i\a}-1,..\ra    &   (17a)  \cr
& a_{i\a}^+ |...,\t_{i\a},..\ra = (1- \t_{i\a})(-1)^{\t_{i1}+..+\t_{i,\a -1}}
 \sqrt{p-\sum_{\b} \t_{i\b}} |...,\t_{i\a}+1,..\ra   &  (17b) \cr
}
$$
Moreover
$$
X_{00}^{(i)}|p;\t_{i1},\t_{i2},...,\t_{in}\ra
=(p-\sum_{\b}\t_{i\b}) |p;\t_{i1},\t_{i2},...,\t_{in}\ra.
 \eqno(18)
$$
From (17) one can compute the action of all the even generators. In particular
the number operator $N_{i\a}$ for particles of flavor $\a$ on the site $i$  is
(see (13a))
$$
N_{i\a}=\{a_{i\a}^+,a_{i\a}^-\}-X_{00}^{(i)}=X_{\a \a}^{(i)},
\eqno(19)
$$
namely,
$$
N_{i\a}|..,\t_{i\a},..\ra= \t_{i\a}|..,\t_{i\a},..\ra. \eqno(20)
$$

The operators $X_{00}^{(i)}$ do not belong to $sl(1|n)$. Nevertheless within
each irreducible module $W(n,p,i)$ these operators can be expressed via the
$sl(1|n)$ generators:
$$
X_{00}^{(i)}={1\over{n-1}}\big(\sum_{\a=1}^n \{a_{\a i}^+,a_{\a i}^-\} -p
\big).\eqno(21)
$$
Then
$$
X_{\a\a}^{(i)}=\{a_{\a i}^+,a_{\a i}^-\}-X_{00}^{(i)},~~~\a=1,...,n. \eqno(22)
$$

From(16) there emerges an important conclusion, which is in fact

\s\n {\it The Pauli principle} for $A-$superstatistics (at site
$i$): if the order of statistics is $p$ then each $\t_{\a i}=0,1$
(fermionic like property: on each orbital there can be no more than
one particle), but in addition each site $i$ can accommodate up to
min($p,n$) particles.

The case $p=1$ corresponds to hardcore fermions. The particles, corresponding
to an arbitrary $p$ could be called hardcore fermions of order $p$.

From a mathematical point of view we have constructed explicitly (back in 1980
[1]) a class of representations, the Fock representations, of each local
$gl(1|n)$, labelled by $p=1,2,..$. The construction is relatively easy and this
is due to the fact that the creation operators anticommute. The similar problem
for parastatistics turned to be very difficult. It was soved very recently, see
[10] [11], more than 50 years after the discovery of Green's parastatistics.

We should mention that apart from in [1] the transformation relations of the
state space under the  actions of the Hubbard generators were written down also
in [12]. The results coincide. There is only a difference in the notation. In
particular our $p$ corresponds to the biggest amount of particles to be
accommodated on a site $i$, whereas $n_0$ in [12], (5) and (6) is
$n_0=p-\sum_{b=1}^N \t_\b$ . An interesting speculation would be to interpret
$n_0$ as a number of particles in a reservoir and to study the related
thermodynamics.

In conclusion, we have shown that locally (at each site) the statistics of
hardcore fermions is $p=1$ $A-$superstatistics and more precisely
$sl(1|n)-$statistics of order one. Similar as for parastatistics the related
CAO's obey triple relations (6). The $A-$superstatistics admits also other
configurations. In particular if the order of statistics is $p$, then each site
can accommodate no more than $p$ particles. 

We have indicated that the Hubbard operators (taken at each site) are
generators of the LS $gl(1|n)$ in the usual sense: they constitute a basis in
$gl(1|n)$ considered as a linear space. Based on the circumstance that within
each irreducible module the CAO's generate $gl(1|n)$, we have expressed the
Hubbard operators via the creation and annihilation operators within any Fock
space (for any $p$).

Finally we mention that the Lie superalgebra $gl(1|n)$ has many other
irreducible representations. They are labelled by $n+1$ numbers [13], [14],
[15]. Hence the same holds for the hardcore fermions and for the Hubbard
operators. Are these generalized hardcore fermions of interest? Do they carry
new physical information? These are questions still to be answered and a
motivation for further investigations.

\bigskip\n
{\bf References}

{\settabs\+[1111] & I. Patera, T. D. Palev, Theoretical interpretation of the
   experiments on the elastic \cr

\bigskip\n
\+ 1.   & T.D. Palev, {\it J. Math. Phys.} {\bf 21} 1293 (1980)\cr

\+ 2.  & T.D. Palev, {\it Rep. Math. Phys.} {\bf 31} 241 (1992)\cr

\+ 3.  & J. Hubbard, {\it Proc. Roy. Soc. A} {\bf 276} 238 (1963)\cr

\+ 4. & M.C. Gutzwiller, {\it Phys. Rev. Lett.} {\bf 10} 3 (1963)\cr

\+ 5. & G. Kotliar and A.E. Ruckenstein {\it Phys. Rev. Lett.}
      {\bf 11} 1362 (1986)\cr

\+ 6. & S.E. Barnes, {\it J. Phys. F} {\bf 6} 1375 (1976)\cr

\+ 7. & P. Coleman, {\it Phys. Rev. B} {\bf 29} 3035 (1984)\cr

\+ 8. & N. Read and  D. Newns, {\it J. Phys. C} {\bf 16} 3273
       (1983)\cr

\+ 9. & G. Jackeli and N.M. Plakida, cond-mat/9909019 \cr

\+ 10. & S. Lievens, N.I. Stoilova and J. Van der Jeugt,
          ~~arXiv:0706.4196v1 [hep-th]\cr

\+ 11. &  N.I. Stoilova and J. Van der Jeugt,
          ~~arXiv:0712.1485v1 [hep-th]\cr

\+ 12. & R. Zeyher and M.L. Kulic, {\it Phys. Rev. } {\bf 53B} 2850
      (1996)\cr
\+ 13. & T.D. Palev, {\it Funct. Anal. Appl.} {\bf 21} 85 (1987)\cr

\+ 14. & T.D. Palev,  {\it J. Math. Phys.} {\bf 29} 2589 (1988); {\bf 30} 1443 (1988)\cr

\+ 15. & R.C. King,  N.I. Stoilova and J. Van der Jeugt,
         hep-th/0602169 and  \cr
\+    & {\it J. Phys.} {\bf A 39}, 5763 (2006)\cr

\end